\documentclass[12pt]{elsart}

\usepackage{epsfig,amsbsy,amssymb}
\usepackage{times}

\begin{document}
\begin{frontmatter}

\rightline{
\vbox{
\hbox{SNUTP-99-018}
\hbox{nucl-th/9905044}
}}

\title{\bf Probing nucleon strangeness structure with $\boldsymbol{\phi}$
electroproduction}

\author[SNU]{Yongseok Oh\thanksref{yoh}},
\author[JINR]{Alexander I. Titov\thanksref{titov}},
\author[NTU]{Shin Nan Yang\thanksref{sny}},
\author[KU]{Toshiyuki Morii\thanksref{tm}}

\address[SNU]{Center for Theoretical Physics, Seoul National University,
              Seoul 151-742, Korea}
\address[JINR]{Bogoliubov Laboratory of Theoretical Physics, JINR,
               141980 Dubna, Russia}
\address[NTU]{Department of Physics, National Taiwan University,
              Taipei, Taiwan 10617}
\address[KU]{Faculty of Human Development, Kobe University,
             3-11 Tsurukabuto, Nada, Kobe 657-8501, Japan}

\thanks[yoh]{E-mail address : {\tt yoh@phya.snu.ac.kr}}
\thanks[titov]{E-mail address : {\tt atitov@thsun1.jinr.ru}}
\thanks[sny]{E-mail address : {\tt snyang@phys.ntu.edu.tw}}
\thanks[tm]{E-mail address : {\tt morii@kobe-u.ac.jp}}

\begin{abstract}

We study the possibility to constrain the hidden strangeness content
of the nucleon by means of the polarization observables in $\phi$
meson electroproduction.
We consider the OZI evading direct knockout mechanism that arises from
the non-vanishing $s \bar s$ sea quark admixture of the nucleon
as well as the background of the dominant diffractive and the
one-boson-exchange processes.
Large sensitivity on the nucleon strangeness are found in several
beam-target and beam-recoil double polarization observables. 
The small $\sqrt{s}$ and $W$ region, which is accesible at some of
the current high-energy electron facilities, is found to be the optimal
energy region for extracting out the OZI evasion process.

\end{abstract}


\end{frontmatter}

\newpage

Production of $\phi$ mesons from nucleon targets has been suggested
as a sensitive probe to study the hidden strangeness of the nucleon.
This is because the $\phi$ is a nearly pure $s\bar s$ state so that
its direct coupling to the nucleon is suppressed by the OZI rule.
However, if there exists a non-vanishing $s\bar s$ sea quark component
in the nucleon, the strange sea quark can contribute to the $\phi$
production via the OZI evasion processes.
Investigation of such processes can then be expected to shed light on
the nucleon strangeness content, if any.
For example, recent studies on the $\phi$ production in $p\bar p$
annihilation at rest indicate a large violation of the OZI rule
\cite{Wied97-Amsl98-Sapo98}, which can be explained with the presence of
an intrinsic $s\bar s$ component in the nucleon since it provides with
additional rearrangement and shakeout diagrams \cite{EKKS95,GFYY97}.%
\footnote{It has been, however, also claimed that this OZI violation
could be understood within modified meson exchange models
\cite{LZL93-LZ96-ML98-BL94} by excluding intrinsic strange sea
quark component in the nucleon wavefunction.}

The $\phi$ meson can also be produced from the nucleon with photons and
electrons.
The dominant process in the electromagnetic production of the $\phi$ comes
from the diffractive production (vector-meson dominance) through the
Pomeron exchange, while the one-boson-exchange mechanism gives corrections
mostly at backward scattering angles.
In addition, the possible hidden $s\bar s$ cluster in the proton can
contribute through direct knockout process.
The knockout process was first estimated with a nonrelativistic harmonic
oscillator quark model \cite{HKW92}.
In Ref. \cite{TYO97}, we employed a relativistic harmonic oscillator quark
model (RHOQM) to include the relativistic Lorentz-contraction correction.
Following the analysis of Ref. \cite{HKW92}, we found that the theoretical
upper bound of the admixture of strange sea quarks in the proton allowed
for by the existing electroproduction cross section data is less than $5$\%.
Nevertheless, it is {\it not easy\/} to discern each process in the
cross section measurements because their respective contributions have
similar dependence on momentum transfer.
We have recently demonstrated \cite{TOY97,TOYM98} that many double
polarization observables in $\phi$ photoproduction could be more useful
tools in investigating the strange sea quark structure in the nucleon.
In this paper, we extend our previous study to the case of $\phi$
electroproduction with longitudinally polarized electron beams.

In $\phi$ electroproduction process, $e+p \to e+p+\phi$, using one-photon
exchange approximation, we define the four momenta of the initial
electron, final electron, virtual photon, initial proton, final proton
and produced vector-meson as $k_e$, $k'_e$, $q$, $p, p'$ and $q_{\phi}$,
respectively.
In the laboratory frame, we write $k_e = (E_e, \boldsymbol{k}_e)$,
$k'_e = (E_e', \boldsymbol{k}'_e)$, $p = (E_p^L, \boldsymbol{p}^L) =
(M_N, \boldsymbol{0})$,
$p' = (E_{p'}^L, \boldsymbol{p'}^L)$, $q_\phi = (\omega_\phi^L,
\boldsymbol{q}^L_{\phi})$
and $q = (\nu^L, \boldsymbol{q}^L)$, where $M_N$ is the nucleon mass.
In the hadronic (or $\gamma^* p$) c.m. frame, we write
$q = (\nu, \boldsymbol{q})$, $q_\phi = (\omega_\phi, \boldsymbol{q}_\phi)$,
$p = (E_p, -\boldsymbol{k})$ and $p' = (E_{p'}, -\boldsymbol{q}_\phi)$,
respectively.
We further define $s = (k+p)^2$, $Q^2 = - q^2$, $W^2 = (p+q)^2$ and
$t = (p-p')^2$.

\begin{figure}[t]
\centering
\epsfig{file=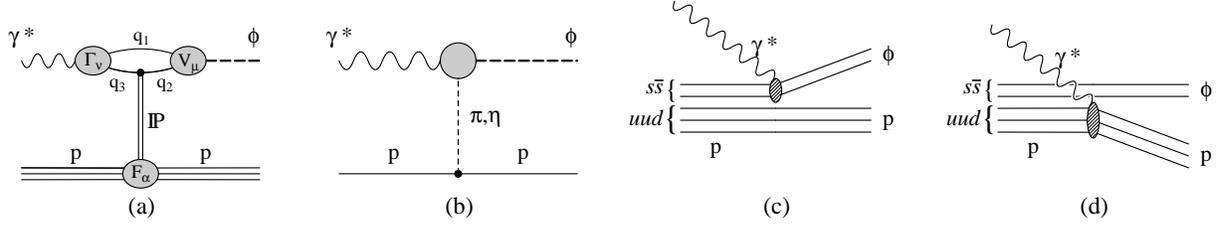, width=\hsize}
\caption{Processes for $\phi$ meson electroproduction. (a) diffractive,
(b) one-boson-exchange, (c) $s\bar s$-knockout and (d) $uud$-knockout
process.}
\label{fig:models}
\end{figure}

Our model for $\phi$ electroproduction includes the diffractive production,
one-boson-exchange ($\pi$ and $\eta$ exchange) and the direct knockout
processes as shown in Fig. \ref{fig:models}.
The OZI evaded knockout process is allowed only if the proton contains
non-vanishing $s\bar s$ sea quark admixture.
In the vector-meson-dominance model (VDM) of the diffractive production
process \cite{CABD81,BSYP78}, the incoming photon first converts into
$q \bar q$ pair ($\phi$-meson in our case) and this $\phi$ scatters
diffractively from the nucleon target through Pomeron exchange as shown in
Fig. \ref{fig:models}(a).
It has been claimed that most of the vector-meson electromagnetic production
process data could be understood qualitatively and quantitatively by the
diffractive process with the Pomeron-photon analogy
\cite{DL84-DL86,PL96-PL97,LM95}.%
\footnote{Here we do not consider the two-gluon exchange model for the
Pomeron exchange \cite{DL87a,LN87,Cud90}.}
As in our previous works on $\phi$ photoproduction \cite{TOY97,TOYM98},
we use the parameterization of the vector-meson-dominance model
\cite{BHKK74} together with the spin structure coming from the Pomeron-photon
analogy. (See Ref. \cite{TOYM98} for more details.)

Then the invariant amplitude of this process can be written as
\begin{equation}
T^{\rm VDM} = i T_0 \varepsilon_\mu^* (\phi) \mathcal{M}^{\mu\nu}
\varepsilon_\nu (\gamma),
\label{T_VDM}
\end{equation}
with $\varepsilon_\mu (\phi)$ and $\varepsilon_\mu (\gamma)$ the $\phi$
and photon polarization vector, respectively, and
\begin{equation}
\mathcal{M}^{\mu\nu} = 
\bar u (p') \gamma_\alpha u(p)\, \left\{
(q + q_\phi)^\alpha g^{\mu\nu}
- 2 q^\mu g^{\alpha\nu}
- \frac{2 q^2} {q \cdot q_\phi} q_\phi^\nu g^{\alpha\mu} \right\},
\end{equation}
where $u(p)$ is the Dirac spinor of the nucleon with momentum $p$ and
we keep the relevant terms only \cite{TOYM98}.

The dynamics of the Pomeron-hadron interactions is represented by $T_0$.
It is parameterized according to the prescription of Ref. \cite{FS69},
which gives the differential cross section for virtual photoproduction as
\begin{equation}
\frac{d\sigma_{\gamma^*}}{dt} =
\frac{\sigma_\phi(0,W)}{(1+Q^2/M_\phi^2)^2}
\frac{q^*(0)}{q^*(Q^2)} ( 1 + \varepsilon R_\phi ) b_\phi \exp \{ b_\phi
[t - t_{\rm max} (0)] \},
\label{dsig_VDM}
\end{equation}
where $\varepsilon$ is the virtual photon polarization parameter,
$M_\phi$ the $\phi$-meson mass and
\begin{equation}
q^*(Q^2) = \frac{1}{2W} \sqrt{\left[ (W - M_N)^2 + Q^2 \right]
\left[ (W + M_N)^2 + Q^2 \right]}.
\end{equation}
The VDM hypothesis leads to that $R_\phi$, the ratio of the cross sections
with longitudinal and transverse photons, is
$R_\phi = \xi^2 Q^2 / M_\phi^2$, where the phenomenological factor
$\xi^2$ is to be determined by experiments.
The parameters are fixed as $\xi^2 = 0.328$ \cite{CABD81,DGHH77-DGLS79},
$\sigma_\phi(0,W) = 0.20 \mbox{ $\mu$b }$ and $b_\phi = 4.01 \mbox{ GeV}^2$
for $W \sim 2$ GeV \cite{BHKK74}, and
$\sigma_\phi(0,W) = 0.22 \mbox{ $\mu$b }$ and $b_\phi = 3.46 \mbox{ GeV}^2$
for $W \le 3$ GeV \cite{CABD81,DGHH77-DGLS79}.
As in photoproduction \cite{TOYM98}, this amplitude is purely imaginary
and has the helicity conserving form at the forward scattering angles,
i.e., at small $|t|$ limit.

The one-boson-exchange (OBE) process shown in Fig. \ref{fig:models}(b)
is allowed because of the possible decays of $\phi \to \gamma\pi$ and
$\phi \to \gamma\eta$.
This process represents the contributions from the possible non-strange
quark component in the $\phi$-meson and gives a correction to the
diffractive production \cite{JLMS77}.
In fact, OBE is comparable to or even dominates the diffractive
process in large $|t|$ region.
In order to calculate OBE process we employ the pseudovector coupling
for $\pi NN$ and $\eta NN$ interactions with the
Gell-man--Sharp--Wagner (GSW) model for $\phi$ decays, i.e., the $\phi$
decays into $\gamma \pi$ ($\gamma\eta$) only through the intermediate
$\rho$ ($\omega$) vector meson.
Then the effective Lagrangian for the $\pi$-exchange reads
\begin{equation}
\mathcal{L} = \frac{g_{\pi NN}}{2M_N} \bar{N} \gamma^\mu \gamma_5
\boldsymbol{\tau} \cdot \partial_\mu \boldsymbol{\pi} N
+ g_{\phi\rho\pi} \epsilon^{\mu\nu\alpha\beta} \partial_\mu \phi_\nu
\,\mbox{Tr}\, ( \partial_\alpha \rho_\beta^0 \pi^0 ).
\end{equation}
The effective Lagrangian for the $\eta$-exchange is obtained by the same
manner.
We use the $\pi NN$ coupling constant $g^2_{\pi NN}/4\pi = 14.0$
\cite{Timm98} and rely on the SU(3) relation for the $\eta NN$ coupling
constant, which gives ${g_{\eta NN}}/{g_{\pi NN}} = 0.26$ using $F/D =
0.565$.
The effective coupling constants of $\phi\gamma\pi$ and
$\phi\gamma\eta$ are related with $g_{\phi\rho\pi}$ and
$g_{\phi\omega\eta}$ and determined from $\phi$ decay widths
and the form factors are used for each vertex in the form
of $(\Lambda^2 - M^2) / (\Lambda^2 - t)$, where $M$ stands for
the corresponding pseudoscalar meson mass \cite{TLT98}.
Note that this OBE amplitude is purely real.

If the nucleon contains any $s\bar s$ sea quark admixture, the incoming
photon can interact with the quark clusters, which gives the direct
knockout mechanisms as shown in Fig. \ref{fig:models}(c,d).
They can be classified into $s\bar s$-knockout and $uud$-knockout
according to the struck quark cluster.
To estimate the knockout contribution to the $\phi$ production, the
proton wavefunction is approximated as
\begin{eqnarray}
|p\rangle = A | [uud]^{1/2} \rangle +
\sum_{j_{s\bar s} = 0,1; \, j_c} b_{j_{s\bar s}}
| \left[ \boldsymbol{[} [uud]^{j_{uud}} \otimes
[\boldsymbol{L}]\boldsymbol{]}^{j_c}
\otimes [s\bar s]^{j_{s\bar s}} \right]^{1/2}
\rangle ,
\label{protonwf}
\end{eqnarray}
where the superscripts $j_{uud}$ ($=1/2$) and $j_{s\bar s}$ ($=0,1$)
denote the spin of the corrsponding cluster and $(b_0, b_1)$ correspond
to the amplitudes of the $s\bar s$ cluster with spin $0$ and $1$,
respectively.
The nucleon strangeness $|B|^2$ is then given by $|B|^2 = |b_0|^2 +
|b_1|^2$ with the constraint $|A|^2+|B|^2=1$.
In order to have the positive parity ground state the orbital
angular momentum between the clusters is constrained to be $\ell = 1$.%
\footnote{Detailed discussions on this form of the proton wavefunction can
be found, e.g., in Refs. \cite{HKW92,TYO97,TOY97,TOYM98}.}
We use the RHOQM for the hadron radial wavefunctions.

In the hadronic c.m. system, the amplitudes of the knockout process can
be expressed as
\begin{eqnarray}
T^{s\bar s}_{m_\phi, m_f; \lambda_\gamma, m_i} &=&
i T_0^{s\bar s} \mathcal{S}^{s\bar s}_{m_\phi, m_f; \lambda_\gamma, m_i},
\nonumber \\
T^{uud}_{m_\phi, m_f; \lambda_\gamma, m_i} &=&
i \left( T_0^{uud} \mathcal{S}^{uud}_{m_\phi, m_f; \lambda_\gamma, m_i}
+  T_1^{uud} \mathcal{Z}^{uud}_{m_\phi, m_f; \lambda_\gamma, m_i} \right),
\label{T-KO}
\end{eqnarray}
where $T_0^{s\bar s}$ and $T_{0,1}^{uud}$ include the energy and momentum 
transfer dependence of the amplitudes. (For their explicit expressions,
see Refs. \cite{TOYM98,OTYM99b}.)
Their spin structures are given as \cite{TYO97,TOY97,TOYM98,OTYM99b}, 
\begin{eqnarray}
\mathcal{S}^{s\bar s}_{m_\phi, m_f; \lambda_\gamma, m_i} &=&
\sqrt{3} \sum_\varrho
\langle {\textstyle\frac12}\, m_f\, 1\, \varrho \, | \,
{\textstyle\frac12}, m_i \rangle \, \xi^{s\bar s}_\varrho \,
\lambda_\gamma \boldsymbol{\varepsilon}_{m_\phi} (\phi) \cdot
\boldsymbol{\varepsilon}_{\lambda_\gamma} (\gamma),
\nonumber \\
\mathcal{S}^{uud}_{m_\phi, m_f; \lambda_\gamma, m_i} &=&
- \sqrt{3} \sum_{j_c,m_c,\varrho} \langle {\textstyle\frac12}\,
  m_f-\lambda_\gamma\, 1\, \varrho \, | \, j_c\, m_c \rangle
\langle j_c \, m_c \, 1\, m_\phi \, | \, {\textstyle\frac12}\, m_i
\rangle \, \left( 1 - \delta_{\lambda_\gamma,0} \right)
\xi^{uud}_\varrho,
\nonumber \\
\mathcal{Z}^{uud}_{m_\phi, m_f; \lambda_\gamma, m_i} &=&
- \sqrt{\frac32} \sum_{j_c,m_c,\varrho} \langle {\textstyle\frac12}\,
  m_f\, 1\, \varrho \, | \, j_c\, m_c \rangle
\langle j_c \, m_c \, 1\, m_\phi \, | \, {\textstyle\frac12}\, m_i
\rangle \, f_\mu \varepsilon^\mu_{\lambda_\gamma} (\gamma) \xi^{uud}_\varrho,
\label{S:KO}
\end{eqnarray}
with
\begin{eqnarray}
f_0 &=& \frac{5}{6} \left( 1 + \frac{E_{p'}^L - \omega_\phi^L}{M_N} +
2 \frac{\boldsymbol{q}_L \cdot \boldsymbol{p}_L'}{E_{p'}^L M_N}  \right),
\nonumber \\
\boldsymbol{f} &=& \frac{5}{3M_N} \left[ - \boldsymbol{q}_{\phi,L} +
\frac{\boldsymbol{q}_L
\cdot \boldsymbol{q}_{\phi,L}}{|\boldsymbol{q}_L|^2} \boldsymbol{q}_L + \frac{\nu_L}{E_{p'}^L}
\left( \boldsymbol{p}_L' - \frac{\boldsymbol{q}_L \cdot
\boldsymbol{p}_L'}{|\boldsymbol{q}_L|^2}
\boldsymbol{q}_L \right) \right] + \frac{f_0 \nu_L}{|\boldsymbol{q}_L|^2}
\boldsymbol{q}_L,
\end{eqnarray}
and
\begin{equation}
\xi^{s\bar s}_{\pm 1} = \pm \frac{1}{\sqrt2} \sin \theta_{p'},
\qquad
\xi^{s\bar s}_0 = \cos\theta_{p'}, \qquad
\xi^{uud}_{\pm 1} = \mp \frac{1}{\sqrt2} \sin \theta_{q_\phi},
\qquad
\xi^{uud}_0 = \cos\theta_{q_\phi},
\end{equation}
where $\theta_\alpha$ is the production angle of particle $\alpha$ in
the $\gamma^* p$ laboratory frame.

Note that $T^{s\bar s} \propto b_0$ and $T^{uud} \propto b_1$ by the
symmetry properties of the wavefunctions.
Since all of these amplitudes are purely imaginary, they lead to 
strong interference with the diffractive process.
In the kinematical region of our interest, $T_1^{uud}\mathcal{Z}^{uud}$ is
suppressed by the $T_0^{uud} \mathcal{S}^{uud}$ term in the $uud$-knockout,
and the $s \bar s$-knockout dominates the $uud$-knockout at small $|t|$
region.
However, the longitudinal photon can contribute only through the $T_1^{uud}
\mathcal{Z}^{uud}$ part.
Having the $T$-matrix, it is straightforward to construct the helicity
amplitudes.

\begin{figure}[t]
\centering
\epsfig{file=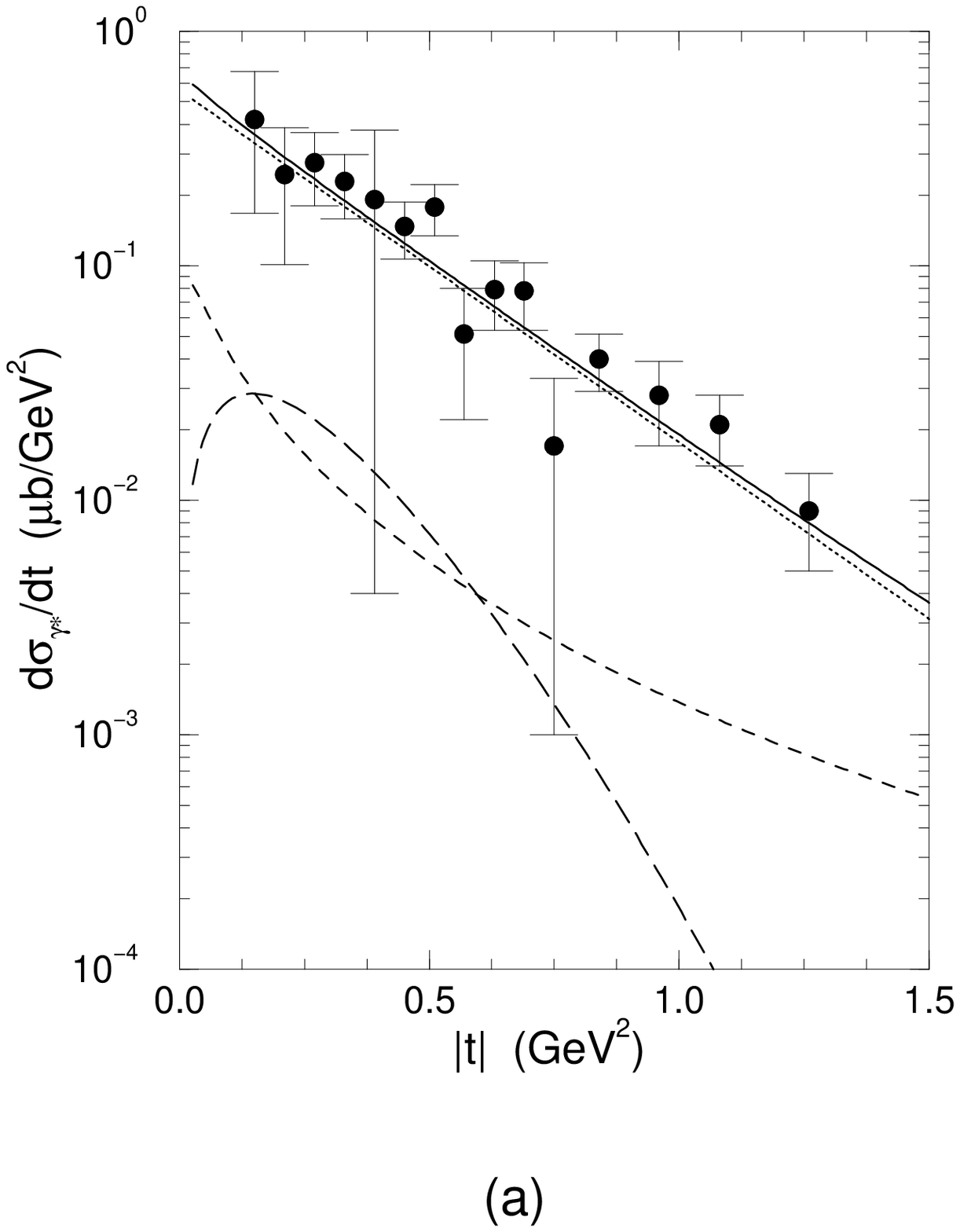, width=5cm} \qquad
\epsfig{file=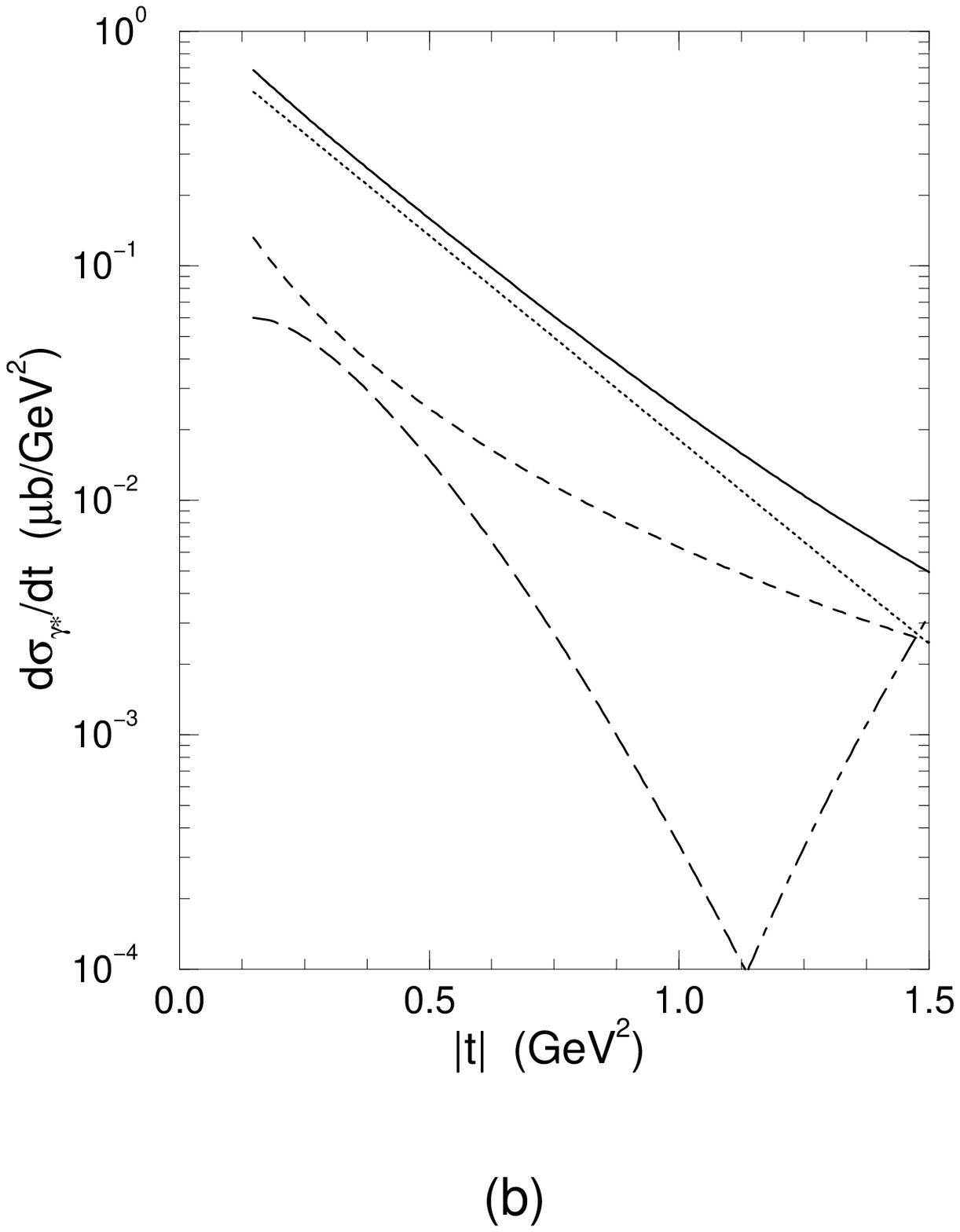, width=5cm}
\caption{The differential cross section for $\gamma^* + p \to \phi + p$
for (a) $\sqrt{s} = 4.73$ GeV, $W = 2.94$ GeV and $Q^2 = 0.23$ GeV$^2$ and
(b) $\sqrt{s} = 2.55$ GeV, $W = 2.15$ GeV and $Q^2 = 0.135$ GeV$^2$.
The dotted, dashed, long-dashed and dot-dashed lines are diffractive
process, OBE, $s\bar s$-knockout and $uud$-knockout processes, respectively,
while the solid lines show the diffractive plus OBE predictions.
In the knockout processes, $|b_0|^2 = |b_1|^2 = 0.5$\% was assumed.
The experimental data are from Ref. \protect\cite{DGHH77-DGLS79}.}
\label{fig:dsigdt}
\end{figure}

We give our predictions on the differential cross sections
$d \sigma_{\gamma^*}/dt$ of virtual photoproduction in
Fig. \ref{fig:dsigdt} at two energy regions:
($\sqrt{s} = 4.73$ GeV, $W = 2.94$ GeV, $Q^2 = 0.23$ GeV$^2$),
where some experimental data exist \cite{DGHH77-DGLS79} and
($\sqrt{s} = 2.55$ GeV, $W = 2.15$ GeV, $Q^2 = 0.135$ GeV$^2$), which is
the energy region of a Jefferson Lab. proposal \cite{BBBC98}.
We see that the contributions from the associate mechanisms have
similar dependence on the momentum transfer $|t|$ and the knockout
process is suppressed by the diffractive process at small $|t|$.
Thus it is very hard to distinguish them from the differential cross
section measurements (in the forward scattering region).

However, as in the case of photoproduction, we find that some double
polarization observables are very sensitive to the hidden nucleon
strangeness.
As typical examples, we consider beam-target and beam-recoil double
asymmetries, where the electron beams are longitudinally polarized.
We define the beam-target asymmetry $A_{\rm BT}^z$ with the target
nucleons polarized along their momentum direction and the beam-recoil
asymmetry $A_{\rm BR}^x$ with the recoil nucleons polarized perperndicular
to the momentum direction but in the scattering plane, which gives
\begin{equation}
A_{\rm BT(BR)} =
\frac{d\sigma(\uparrow\downarrow) - d\sigma(\uparrow\uparrow)}
{d\sigma(\uparrow\downarrow) + d\sigma(\uparrow\uparrow)},
\end{equation}
where the arrows represent the helicities of the incoming electrons and
target (or recoiled) protons.
Note that these polarization observables are defined for $\Theta = 0$
where $\Theta$ is the angle between the normals to the electron scattering
plane and the hadron production plane.

\begin{figure}[t]
\centering
\epsfig{file=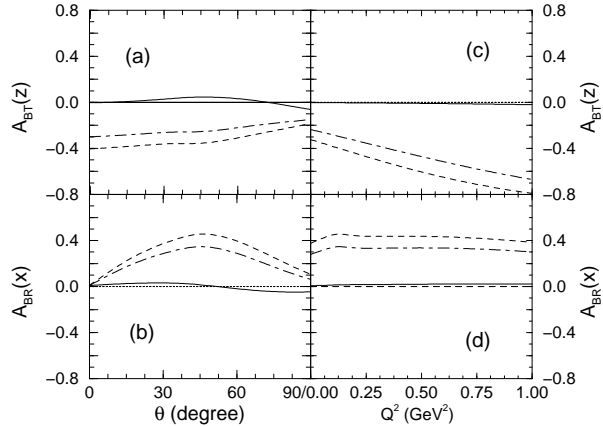, width=8cm}
\caption{Double polarization asymmetries with $\sqrt{s} = 2.55$ GeV and
$W = 2.15$ GeV. (a) $A_{\rm BT}^z (\theta)$ at $Q^2 = 0.135$ GeV$^2$,
(b) $A_{\rm BR}^x (\theta)$ at $Q^2 = 0.135$ GeV$^2$,
(c) $A_{\rm BT}^z (Q^2)$ at $\theta = 0^\circ$
and (d) $A_{\rm BR}^x (Q^2)$ at $\theta = 45^\circ$, where $\theta$ is the
scattering angle in the hadronic c.m. system.
The solid, dot-dashed and dashed lines correspond to $|b_1|^2 (= |b_2|^2)
= 0$, $0.25$\% and $0.5$\%, respectively. The phases of $b_{0,1}$ are
chosen to be $+1$.}
\label{fig:pols}
\end{figure}

The results are given in Fig. \ref{fig:pols}, where the solid, dot-dashed
and dashed lines are the results with $|B|^2 = 0$, $0.5$\% and
$1.0$\%, respectively, where $|B|^2 \equiv |b_1|^2 + |b_2|^2$.
In this paper, we focus on the forward scattering regions since the
backward scattering regions may require not a little modifications on our
models.
In Fig. \ref{fig:pols}(a,b), we give the $t$-dependence of the observables
$A_{\rm BT}^z$ and $A_{\rm BR}^x$, while Fig. \ref{fig:pols}(c,d) show
their $Q^2$ dependence at a given scattering angle in the $\gamma^* p$
c.m. system.
Our observations indicate that these double polarization observables in
$\phi$ electroproduction are sensitive to the hidden nucleon strangeness
and can be useful in investigating hidden nucleon structure.
Even with less than $1.0$\% admixture of the nucleon strangeness, the
deviations from the predictions without nucleon strangeness at forward
scattering angles can be large enough to be detected by experiments.
This is because the associated mechanisms have different spin structure.
Figure \ref{fig:pols}(c,d) show that such deviation becomes larger
for $A_{\rm BT}^z$ and is nearly flat for $A_{\rm BR}^x$ as $Q^2$
increases.
Since the deviation decreases with increasing initial energies
$\sqrt{s}$ and $W$, the optimal energy region to observe this deviation
would be near threshold.

Since our results depend on the models for VDM and nucleon wave
function, we have several comments on the possible modifications on our
models.
For example, there may be some corrections to the double polarization
observables by the complex nature of the Pomeron exchange amplitude, which
can interfere with the OBE part.
The recent estimation of Ref. \cite{TLT98} shows that in the beam-target
asymmetry of $\phi$-photoproduction this interference is comparable to the
effect of the knockout process with $|B|^2$ at the level of $0.1$\% at
$\theta \to 0$ limit.
So our results should be understood with this error range.
Since the double asymmetries depends sensitively on the hidden nucleon
strangeness at small energies, it would be also natural to see how much the
final state interactions could change our results.
Also the effect of direct $\phi NN$ coupling, although expected to be
small, and the role of intermediate hadron states might be considered for
large scattering angle regions.

As a conclusion, we find that measurements of the double polarization
observables in $\phi$ electroproduction may provide us with useful
information on the nucleon structure, especially the hidden nucleon
strangeness.
It will be, therefore, very interesting if experiments of this kind can
be carried out at current high-energy electron facilities.

This work was supported in part by
the KOSEF through the CTP of Seoul National University,
the Russian Foundation for Basic Research under grant No. 96-15-96423,
the National Science Council of ROC under grant No. NSC88-2112-M002-015
and Monbusho's Special Program for Promoting Advanced Study (1996, Japan).
Y.O. is grateful to the Physics Department and the Center for Theoretical
Sciences of the National Taiwan University for the warm hospitality.

\rule{\textwidth}{0.7mm}

\end{document}